# A Taxonomy to Assess and Tailor Risk-based Testing in Recent Testing Standards

**Jürgen Großmann,** *Fraunhofer FOKUS, Germany*

**Michael Felderer,** *University of Innsbruck, Austria & Blekinge Institute of Technology, Sweden*

**Johannes Viehmann,** *Fraunhofer FOKUS, Germany*

**Ina Schieferdecker,** *Fraunhofer FOKUS & TU Berlin, Germany*

*This article provides a taxonomy for risk-based testing that serves as a tool to define, tailor, or assess risk-based testing approaches in general and to instantiate risk-based testing approaches for the current testing standards ISO/IEC/IEEE 29119, ETSI EG and OWASP Security Testing Guide in particular. We demonstrate the usefulness of the taxonomy by applying it to the aforementioned standards as well as to the risk-based testing approaches SmartTesting, RACOMAT, PRISMA and risk-based test case prioritization using fuzzy expert systems. In this setting, the taxonomy is used to systematically identify deviations between the standards' requirements and the individual testing approaches so that we are able to position and compare the testing approaches and discuss their potential for practical application.*

## Introduction

Since the systematic integration of risk assessment and testing is a relevant approach to address product risks in software development and to cope with limited testing resources, current standards like ISO/IEC/IEEE 29119, ETSI EG 203 251 or the OWASP Security Testing Guide recommend a systematic integration between these two domains. The systematic combination of risk assessment and testing is known as risk-based testing and it applies assessed risks of the software product as the guiding factor to steer all phases of a test process, i.e., test planning, design, implementation, execution, and evaluation [1]. Risk-based testing has become quite popular and several approaches were developed (see Erdogan et al. [2] for a comprehensive survey of risk-based testing approaches). However, the standards stay mostly abstract with regard to the concrete implementation and do often not provide concrete guidance how to define, adapt, or assess risk-based testing approaches and tools. Because of the growing demand on risk-based testing processes by industry and the increasing number of available risk-based testing approaches, a solid methodological support to define, tailor, categorize, assess, compare, and select risk-based testing approaches is required. The taxonomy presented in this article provides this kind of methodological support. We demonstrate its power and value by applying the taxonomy, which comprises the top-level categories *context*, *risk assessment* and *risk-based test strategy*, to the requirements coming from recent testing standards and relating these requirements to four current risk-based testing approaches. Especially practitioners get a systematic overview of the requirements from standardization, by which techniques and procedures these requirements can be instantiated, and how risk-based testing approaches can be tailored and compared.

## Current Testing Standards

In this section we provide a short overview of the current testing standards ISO/IEC/IEEE 29119, ETSI EG 203251, and OWASP Testing Guide in general and how they address the integration of risk assessment and testing in particular. Later, these standards will be used as the basis to develop a taxonomy for risk-based.

### ISO/IEC/IEEE 29119

The new international series of software testing standards ISO/IEC/IEEE 29119 consists of five parts, which cover (1) Concepts & Definitions, (2) Test Processes, (3) Test Documentation, (4) Test Techniques, as well as (5) Keyword Driven Testing. ISO/IEC/IEEE 29119 explicitly specifies risk considerations to be an integral part of the test planning process. The second part of ISO/IEC/IEEE 29119-2:2013 [3] on test processes follows a risk-based testing process. The test planning process defined by ISO/IEC/IEEE 29119-2:2013 is used to develop a test plan. Its sequence of activities is shown in Figure 1. The highlighted activities 'Understand Context', 'Identify & Analyse Risks', 'Identify Risk Mitigation Approaches', and 'Design Test Strategy' are explicitly risk oriented. To understand the context, it is recommended to consider a project risk register for information on identified project and product risks. During risk identification and analysis, previously identified risks are reviewed, additional risks that can be addressed by software testing are identified, and then risks are estimated and assigned to risk levels to complete risk assessment. During the identification of risk mitigation approaches appropriate means of treating the risks are identified, which are the basis for the definition of a risk-based test strategy during test strategy design.

### ETSI EG 203 251

EG 203251 [4] is an ETSI document that introduces a set of methodologies that integrate security risk assessment and security testing in a systematic manner. This includes both, risk assessment aimed to improve security testing and test activities used to improve the security risk assessment. The guide details how results from security testing can improve risk assessment by providing feedback on existence and distribution of actually existing vulnerabilities. In this scenario, risk values (e.g., likelihood estimates) are adjusted on basis of the test results. Moreover, the guide shows how risk assessment results are used to guide and focus the testing process by identifying the areas of risk within the target's business processes and building and prioritizing the testing program around these risks. In this setting the notion of risk helps focusing the testing resources on the areas that are most likely to cause concern. Moreover, it supports the selection of adequate test techniques on basis of already identified and known threat scenarios. The activities and their level of specification refer to standards like ISO 31000 [5] and ISO 29119 so that they are applicable for a larger number of security testing and risk assessment processes at hand.

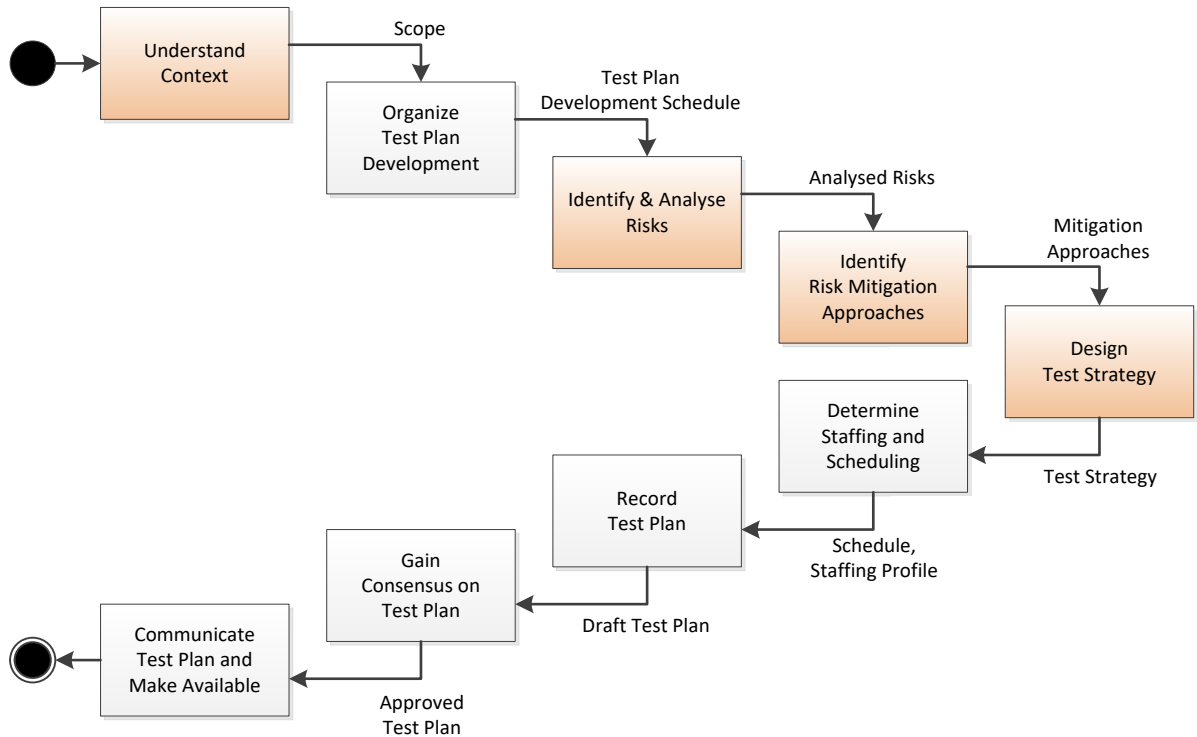

Figure 1. Test Planning Process defined in ISO/IEC/IEEE 29119-2:2013 [3]. The activities 'Understand Context', 'Identify & Analyse Risks', 'Identify Risk Mitigation Approaches', and 'Design Test Strategy' which are explicitly risk oriented are highlighted.

**OWASP Testing Guide**

The OWASP Testing Guide [6] has been developed by the OWASP community and focuses primarily on web application security testing. The guide is a detailed description of the various kinds of testing that is required in a web application security testing process. It describes testing methods and related activities ranging from early phases in software development process (SDLC) until maintenance and operation. Beside others, risk assessment and the notion of risk is addressed as an explicit means to drive the allocation of testing resources, to identify and prioritize testing requirements and to identify adequate testing techniques and test data. Moreover, risk management is used to put test results in context and help identifying the technical, regulatory and business impact of findings and vulnerabilities that are discovered during testing. The OWASP guide recommend the use of predefined risk template for already known security threats and refers to security risk assessment standards like NIST 800-30 [7].

## Taxonomy of Risk-Based Testing

The taxonomy of risk-based testing is shown in Figure 2. It refines a previously published taxonomy for risk-based testing [8] and comprises the top-level classes *context*, *risk assessment* as well as *risk-based test strategy*. It helps to define, adapt, or assess risk-based testing approach according to the requirements coming from recent industrial standards.

**Context**

The *Context* characterizes the overall context of the risk assessment and testing process. It includes the subclasses *Risk driver* to characterize the drivers that determine the major assets, *Quality* *property* for the overall quality objectives that need to be fulfilled and *Risk item* for the elements that are subject to evaluation by risk assessment and testing.

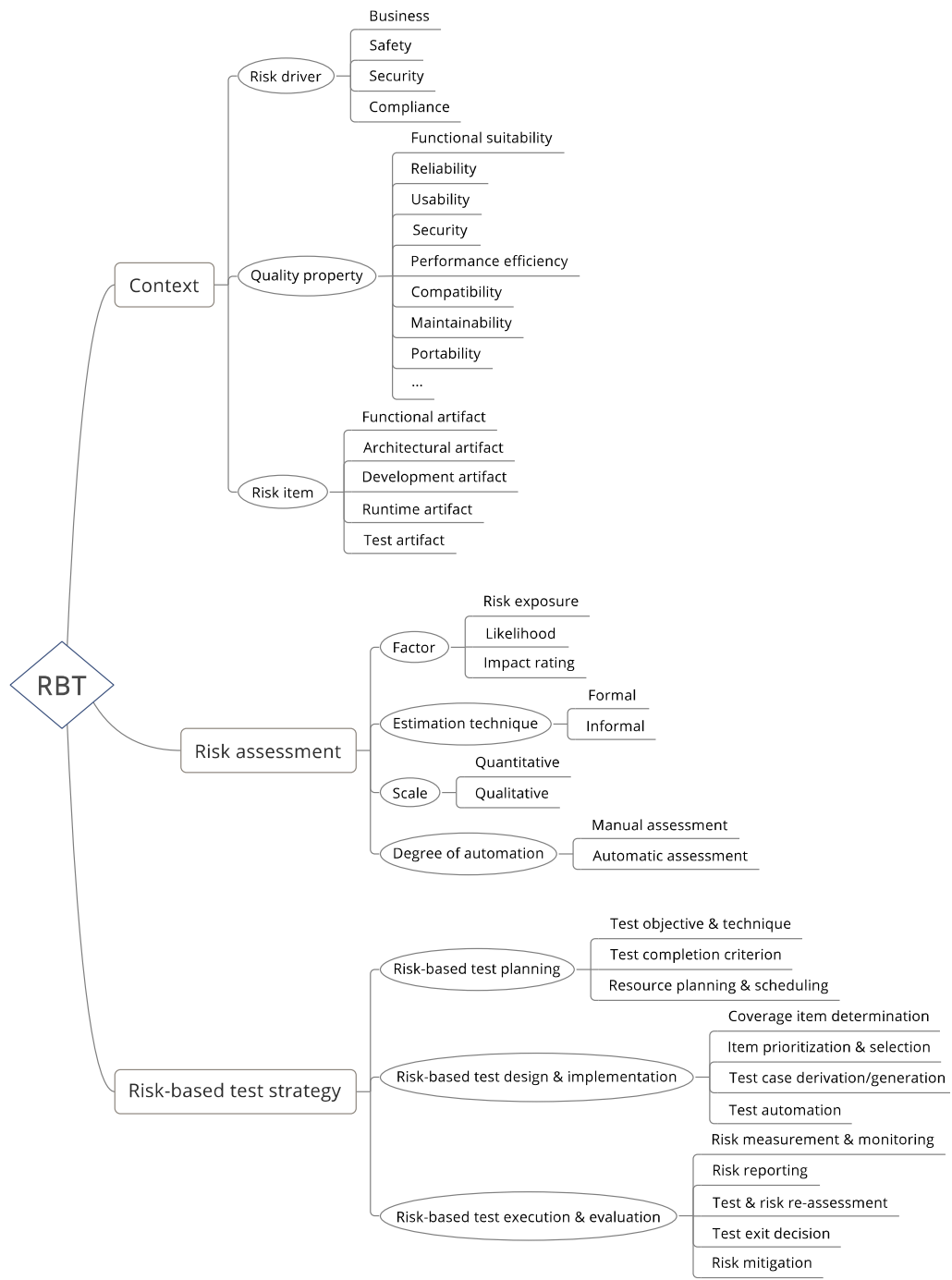

Figure 2. Risk-Based Testing Taxonomy.

**Risk Assessment**
*Risk assessment* characterizes how risks are being determined. Risks are typically expressed as the combination of the consequences of events and the associated likelihoods of occurrence. Risk assessment itself is further differentiated into several classes: *Factor* defines the characteristics determining risk (i.e., risk exposure, likelihood, impact rating); *Estimation technique* determines whether the risk is determined formally based on a model or informally based on simple lists; *Scale* to determine whether quantitative numeric or qualitative ordinal values like low, medium, or high are used for risk values; *Degree of automation* finally determines whether the assessment is done manually or automated.

**Risk-Based Testing Strategy**
Based on the risks being determined and characterized, risk-based testing follows the fundamental test process or variations thereof. The notion of risk can be used to optimize already existing testing activities by introducing a *Risk-based strategy* for prioritization, automation, selection, resource planning etc. This taxonomy aims for highlighting and characterizing the specifics of risk-based testing by relating the activities to the major phases of a normal test process, which can be grouped into three areas: *Risk-based test planning*, *Risk-based test design & implementation*, we well as *Risk-based test execution & evaluation* each with the sub-issues shown in the lower part of Figure 2.

# Current Approaches to Risk-Based Testing

In this section, we present four current approaches to risk-based testing, which we than classify according to the presented taxonomy of risk-based testing. Two of the risk-based testing approaches have already a longer history in industry, i.e. the PRISMA approach, or in academia, i.e. risk-based test case prioritization using fuzzy expert systems, respectively. Two other approaches, i.e., the SmartTesting and the RACOMAT approach, result from our own applied research together with industry partners.

**The SmartTesting Approach for Risk-Based Test Strategy Development**
SmartTesting [9] provides a process for risk-based test strategy development that has been created and evaluated in close collaboration with industry [9,10]. The SmartTesting process consists of seven core steps, i.e., (1) definition of risk items, (2) probability estimation, (3) impact estimation, (4) computation of risk values, (5) determination of risk levels, (6) definition of test strategy, and (7) refinement of test strategy, as well as defect management, requirements management and quality management which are used to establish the preconditions for the process by linking test strategy development to the related processes.

In the SmartTesting approach, first risk items are identified, which are typically derived from the functional structure of the software system, but they can also represent non-functional aspects or system properties. Then, for each risk item a probability value as well as an impact value is derived. The probability value, which expresses the likelihood of defectiveness, often relies on historical defect data. The impact value, which expresses the consequences of risk items being defective, is estimated and typically closely related to the severity values usually determined in requirements engineering. Then overall risk values are computed based on the probability and impact estimates, which are then partitioned into risk levels. Finally, the test strategy is defined and refined on the basis of the different risk levels. For each risk level the test strategy describes how testing is organized and performed.

**The RACOMAT Approach**
RACOMAT [11] is a tool for risk management according to the ISO 31000 standard that has been initially developed in the RASEN research project (http://www.rasenproject.eu). The tool combines component based compositional security risk assessment with automated security testing in the context of risk drivers like security, business and safety. For formally modelling and assessing risks, RACOMAT uses an extended version of CORAS [12] and associates risk analysis artefacts with system models in order to enable automated testing. During the initial risk identification, RACOMAT can automatically generate system models and it suggests common weaknesses and threat scenarios which are most relevant for the analyzed systems. Taking benefit of libraries containing risk assessment artefacts (e.g. Mitre CWE and CAPEC catalogues) and of libraries containing testing artefacts like security test patterns and test metrics, the RACOMAT tool offers a high level of reusability. RACOMAT allows to select and prioritize the not yet tested elements having the greatest impact on the overall risk picture or having the most uncertain likelihood estimates so far.

Besides supporting highly automated risk-based security testing, the RACOMAT tool also supports test based risk assessment. It integrates models approximating the behavior of the SUT observed as security testing results tightly into risk and business models. RACOMAT then performs event simulations with these approximating models to analyze likelihoods and consequences [13]. Calculating overall risk values as expected costs per time period, it assists managers and stakeholders to decide if they accept the risks or if risk treatment is required.

The more accurate risk model updated with the help of test results can be used to start another round of risk-based security testing in an iterative, adapting process.

**The PRISMA Approach**
The Product RISk MAnagement (PRISMA) approach [14] starts with the systematic identification of software product risks. It distinguishes business risks originating from the most important parts of the product, e.g., areas with critical functionalities, visible areas, or most used areas as well as technical risks with a high defect probability, e.g., complex areas, areas with a lot changes, areas with a high degree of collaboration. These criteria are weighted to calculate the overall risk of the risk items. PRISMA aims for identifying these risks, evaluating and visualizing them, and defining a differentiated test approach for differentiable risks.

The central artifact of the PRISMA method is a so called Product Risk Matrix. It consists of four quadrants each representing a different risk level and motivating a different test approach. Individual risk items, i.e. the items to be tested, can be placed in the matrix dependent on a set of predefined risk factors.

From a process point of view, PRISMA starts with an overall planning phase. addressing the identification of the risk items, the definition of the impact and likelihood factors, and the identification of stakeholders for the risk items. After the planning phase, PRISMA envisages an optional kick-off meeting, followed by iterative steps to collect, validate and coordinate the assignment and scoring of risk factors for each risk item among different stakeholders. During this process, the final scores are determined and represented in the final version of the risk matrix. In a last step, the risk items are prioritized and associated with a differentiated test approach based on their position in the risk matrix. Such differentiated test approaches may vary with respect to test depth or priorities in testing. For instance, a test approach for a high risk area may have more reviews, more test cases, better coverage, stricter exit criteria, or just more experienced testers.

The PRISMA method provides basic tool support and is scalable. For larger projects, the approach envisions multiple risk matrices so that a larger set of risk items can be maintained. The PRISMA approach has been used on all testing levels, i.e., for component, integration, system and acceptance testing, and applies to all quality attributes (i.e., functional suitability, reliability, usability, security, etc.) that are testable. It has been applied in many projects and companies covering different industries.

**Risk-Based Test Case Prioritization Using Fuzzy Expert Systems**

The overall goal of Risk-Based Test Case Prioritization Using Fuzzy Expert Systems [15] is to support the prioritization of requirements-based tests by making the requirements risk estimation process more systematic, precise and less subjective by using fuzzy expert systems. The approach comprises four steps, (1) risk estimation by correlating with requirements, (2) risk exposure calculation for requirements, (3) risk exposure calculation for risk items, and (4) prioritization of requirements and test cases.

In the first step, the risk indicators requirements complexity, requirements size, requirements modification status and potential security threats, that have been proven to effectively indicate defects in software systems, are determined. On the one hand, requirements complexity and size are objectively determined based on source code and requirements information. On the other hand, requirements modification status and potential security threats can be subjective and are therefore determined based on requirements information utilizing a fuzzy expert system to reduce the subjectivity and possible errors made by human judgement.

In the second step, the risk exposure of a requirement is calculated as the weighted mean of the risk indicator values for that requirement. The weights are calculated based on the analytic hierarchy process. In the third step, the risk exposure values for risk items are calculated by summing up the products of risk exposure values of requirements associated with the risk item and their severity values.

Finally, in the last step requirements and test cases related to the requirements are prioritized based on risk exposure values for requirements derived from the risk exposure values of risk items linked to the requirements. The prioritization of test cases is explicitly applied for risk-based regression test prioritization and selection to improve the fault detection rate and to find more faults in risky components earlier. However, the risk exposures assigned to requirements and risk items can also be used to steer other test activities like test case design or automation as well.

## Application of the Taxonomy

Table 1 applies and evaluates the taxonomy of risk-based testing by using it as a systematic tool to relate the requirements and recommendations from standards to the capabilities of the approaches and tools described above. The taxonomy helps testers and managers to understand in which areas risk-based testing activities are applicable in general and in which areas standards already recommend activities. Applying the taxonomy to a set of existing approaches, we additionally evaluate the coverage of standard requirements by approach and compare the approaches with each other considering their area of application and their capabilities. Since the standards and approaches are not explicit with respect to several differentiations that are used in the, we use the following notation in Table 1.

Symbols for Standards:
*X:*     *explicitly mentioned and recommended*
*O:*     *differentiation not made but reasonable*
*An empty cell means not covered by the standard*

Symbols for approaches:
*++:     fully supported or elaborated by the approach*
*+:      partially supported or elaborated by the approach*
*(T):    an additional (T) means that the approach provides dedicated tool support for the respective area*

*An empty cell means not supported or elaborated by the approach*

**Table 1. Classification of the SmartTesting approach and the RACOMAT tool to risk-based testing according to the taxonomy of risk-based testing.**

|  |  |  | *Standards* | | | *Approaches* | | | |
|---|---|---|---|---|---|---|---|---|---|
|  |  |  | ISO 29119 | ETSI EG | OWASP Testing Guide | SmartTesting | RACOMAT | PRISMA | Prioritization using Fuzzy Expert System |
| Context | | | | | | | | | |
| | Risk driver | | | | | | | | |
| | | Business | X | X | X | ++ | ++ | ++ | + |
| | | Safety | O | | | + | + | + | ++ |
| | | Security | O | X | X | + | ++ | + | ++ |
| | | Compliance | X | | X | + | + | + | + |
| | Quality property | | | | | | | | |
| | | Functional suitability | O | O | O | ++ | + | ++ | ++ |
| | | Reliability | O | O | X | + | ++ | + | |
| | | Usability | O | | O | + | | + | |
| | | Security | O | X | X | + | ++ | + | ++ |
| | | Performance efficiency | O | O | O | + | + | + | |
| | | Compatibility | O | O | O | + | | + | |
| | | Maintainability | O | | | + | | + | ++ |
| | | Portability | O | | | + | | + | |
| | Risk item | | | | | | | | |
| | | Functional artifact | O | | | ++ | ++ | ++ | ++ |
| | | Architectural artifact | O | X | X | ++ | ++ | ++ | ++ |
| | | Development artifact | O | X | X | + | ++ | ++ | ++ |
| | | Runtime artifact | O | X | X | + | + | ++ | |
| | | Test artifact | | | | + | + | + | + |
| Risk assessment | | | | | | | | | |
| | Factor | | | | | | | | |
| | | Risk exposure | X | O | X | ++ (T) | ++ (T) | ++ (T) | ++ |
| | | Likelihood | X | O | X | ++ (T) | ++ (T) | ++ (T) | ++ |
| | | Impact rating | X | O | X | ++ (T) | ++ (T) | ++ (T) | ++ |
| | Estimation technique | | | | | | | | |
| | | Formal | O | O | O | ++ (T) | ++ (T) | + | ++ |
| | | Informal | O | O | O | ++ (T) | ++ (T) | ++ (T) | |
| | Scale | | | | | | | | |
| | | Quantitative | O | O | O | + (T) | ++ (T) | | ++ |
| | | Qualitative | O | O | O | ++ (T) | + (T) | ++ (T) | + |
| | Degree of automation | | | | | | | | |

|  |  | Standards | | | Approaches | | | |
| --- | --- | --- | --- | --- | --- | --- | --- | --- |
|  |  | ISO 29119 | ETSI EG | OWASP Testing Guide | SmartTesting | RACOMAT | PRISMA | Prioritization using Fuzzy Expert System |
|  | Manual assessment | O | O | O | ++ (T) | + (T) | ++ (T) | + |
|  | Automatic assessment | O | O | O |  | ++ (T) |  | ++ |
| Risk-based test strategy | | | | | | | | |
| Risk-based test planning | | | | | | | | |
|  | Test objective & technique | X | X | X | ++ | + | + |  |
|  | Test completion criterion | X | X |  | + | + | ++ |  |
|  | Resource planning & scheduling | O | X | X | ++ (T) | ++ (T) | ++ (T) | ++ |
| Risk-based test design & implementation | | | | | | | | |
|  | Coverage item determination | O | X | O | + | + | + |  |
|  | Item prioritization & selection | X | X | X | ++ (T) | ++ (T) | ++ (T) |  |
|  | Test case derivation/generation | O | X | X |  | + (T) |  |  |
|  | Test automation | O | O | O |  | ++ (T) |  |  |
| Risk-based test execution & evaluation | | | | | | | | |
|  | Risk measurement & monitoring | X | X | X | + | + (T) |  |  |
|  | Risk reporting | X | X | X | + | ++ (T) | ++ |  |
|  | Test & risk re-assessment | X | X |  | + | ++ (T) | + |  |
|  | Test exit decision | X | X |  | + | + | ++ |  |
|  | Risk mitigation | X | X | X | + | + |  |  |

## Conclusion

Risk-based testing is a powerful technique that helps identifying and testing the relevant parts and properties of a software system, hence detecting critical faults early. Current test standards increasingly recommend risk-based testing. In this article, we have carefully examined the requirements to integrate testing and risk assessment coming from three test standards from relevant bodies like ISO, ETSI and OWASP. We have used our taxonomy of risk-based testing to systematically describe in which areas of testing which of the standards recommend risk-based testing activities and techniques. In addition, we have shown how approaches that are already established or come from our own research, meet the requirements from standards and where they go beyond these recommendation. Considering at least the approaches that have been subject to our evaluation, we can state, that there is already a good coverage of the requirements. However, there are differences between the approaches and not all required areas of risk-based testing are fully supported. Finally, scalable tool support is still missing. Even if methods like RACOMAT, PRISMA and SmartTesting already provide tool support in certain areas, this is not sufficient to cover the need for automation. The RACOMAT tool is still an academic tool and the tool support for SmartTesting and PRISMA are lightweight Excel tools that scale only to a limited extend.

We are aware, that our overall evaluation is not complete. There are much more approaches that address the idea of risk-based testing than we could integrate in our evaluation. However, we see none of them going beyond the capabilities and tools support of the ones that we have examined in this article. Moreover, our approach can be easily extended and applied to additional methods when required. All in all, using the taxonomy has proved to be a good means to systematically analyse and represent the requirements in the area of risk-based testing and a good support in comparing standards and methods.

## About the authors

**Jürgen Großmann** is project manager at Fraunhofer FOKUS. Contact him at juergen.grossmann@fokus.fraunhofer.de.

**Michael Felderer** is a professor at the University of Innsbruck, Austria and a guest professor at the Blekinge Institute of Technology, Sweden. Contact him at michael.felderer@uibk.ac.at.

**Ina Schieferdecker** is director of Fraunhofer FOKUS and professor at the Technical University of Berlin. Contact her at ina.schieferdecker@fokus.fraunhofer.de.

**Johannes Viehmann** is a senior researcher at Fraunhofer FOKUS. Contact him at johannes.viehmann@fokus.fraunhofer.de.